\documentclass[doublecol]{epl2} 

\usepackage{color}

\definecolor{red}{rgb}{1,0,0}
\definecolor{green}{rgb}{0,1,0}
\definecolor{blue}{rgb}{0,0,1}

\title{Superdiffusion of massive particles induced by multi-scale velocity fields}

\author{Reza M. Baram \inst{1,2} \and Pedro G. Lind \inst{1} \and Jos\'e S. Andrade Jr. \inst{2,3} \and Hans J. Herrmann \inst{2,3}}
\institute{                    
\inst{1} CFTC, Universidade de Lisboa, Av. Prof. Gama Pinto 2, 1649-003 Lisbon, Portugal\\
\inst{2} Computational Physics, IfB, HIF E12, ETH H\"onggerberg, CH-8093 Z\"urich, Switzerland\\
\inst{3} Departamento de F\'{\i}sica, Universidade Federal do Cear\'a, 60451-970 Fortaleza, Cear\'a, Brazil
}


\abstract{
We study drag-induced diffusion of massive particles in scale-free velocity 
fields, where superdiffusive behavior emerges due to the scale-free size 
distribution of the vortices of the underlying velocity field. 
The results show qualitative resemblance to what is observed in fluid systems,
namely the diffusive exponent for the mean square separation of pairs
of particles and the preferential concentration of the particles, 
both as a function of the response time.
} 

\begin{document}
\maketitle

\section{Introduction}

Understanding how complex flows advect massive particles is a
challenging problem not only when addressing natural phenomena like
rain drop formation \cite{ref1} and protoplanetary disks \cite{ref2},
but also for environmental and industrial applications such as
dispersion of pollutants \cite{ref3} and design of the combustion
chamber in engines \cite{ref4}.

During the advection of massive particles by the flow, different
transport mechanisms come into play such as the dissipative dynamics
of trajectory attractors and the ejection of the particles from the
vortices (eddies) by centrifugal forces. The interplay of these
mechanisms, when driven by the multi-scale vortical dynamics of
turbulent flows, leads to classical empirical features such as
superdiffusion \cite{zaslavsky02} and preferential concentration (PC)
\cite{prefconEaton1994}.

Experimental and computational studies show that advection of
particles by turbulent flows is generally a superdiffusive process,
that is, the mean squared separation of the particles $\langle
R^2\rangle $ grows faster than $t$, the first power of time \cite{monin75}. 
An early empirical study by Richardson \cite{richardson1926} suggested that 
the mean squared separation of elements of carrier fluid itself grows with $t^3$.

However, preferential concentrations only occur when the density of
particles differs from that of the carrier flow.  The regions of high
vorticity act on the heavier particles like centrifuges pushing them
toward regions of low vorticity and high strains while trapping
lighter particles \cite{prefconEaton1994} which leads to clustering of
particles and inhomogeneity in their distribution.

Experimental difficulties in measuring Lagrangian quantities and limitations of the 
numerical simulations in resolving a wide range of scales of flow motion have posed 
major obstacles in study of these processes \cite{falkovich01}.
During the last decades, great progresses, experimental and computational, 
have been made in overcoming these obstacles (see Ref. \cite{Toschi2009, Salazar2009} for two recent reviews).  

Attempts have been made in using toy models to separate and study the
mechanisms present in real turbulent flows. Bec {\it et al.}
\cite{becPhenomModel2007} introduced a simple flow model to study the
role of ejection of particles by eddies in the distribution of massive
particles. The model consists of a grid of two-dimensional cells with
high vorticities which eject particles with a rate depending on the
response time of the particles to the flow. Their results for the mass
probability distribution in the fully mixed state show qualitative agreement with what is
observed in real systems. However, they do not address the problem of existence of superdiffusive regimes. Moreover, the model lacks a multi-scale
vortical structure, a dominant feature in real turbulent flows, which indeed turns out to be responsible for superdiffusive behaviors.

In this Letter, we introduce a heuristic model for two dimensional
scale-free velocity fields which incorporates a multi-scale vortical structure, 
to study the role of ejection of particles by eddies, 
separated from other mechanisms such as dissipation.  Here, we 
show that the multi-scale structure is a key feature responsible for superdiffusion
regimes in the transport of massive particles. In addition, within the limitations of the model, the results 
obtained show qualitative resemblance to what is observed in real systems. In particular, we 
discuss the applicability of the results to the problem of preferential concentration and Richardson's law. 

\begin{figure}[t]
\begin{center}
\includegraphics*[width=0.24\textwidth]{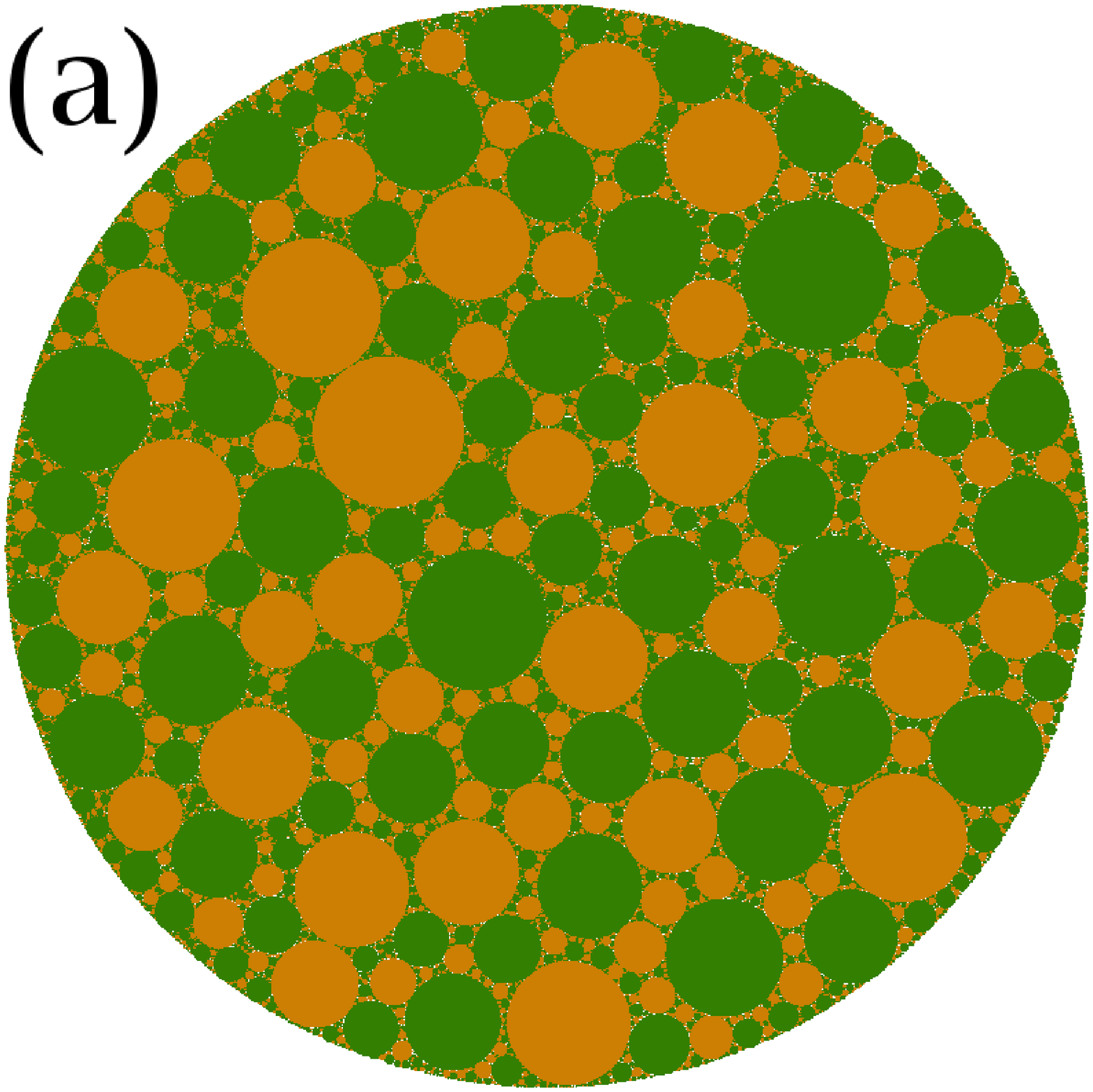}%
\includegraphics*[width=0.24\textwidth]{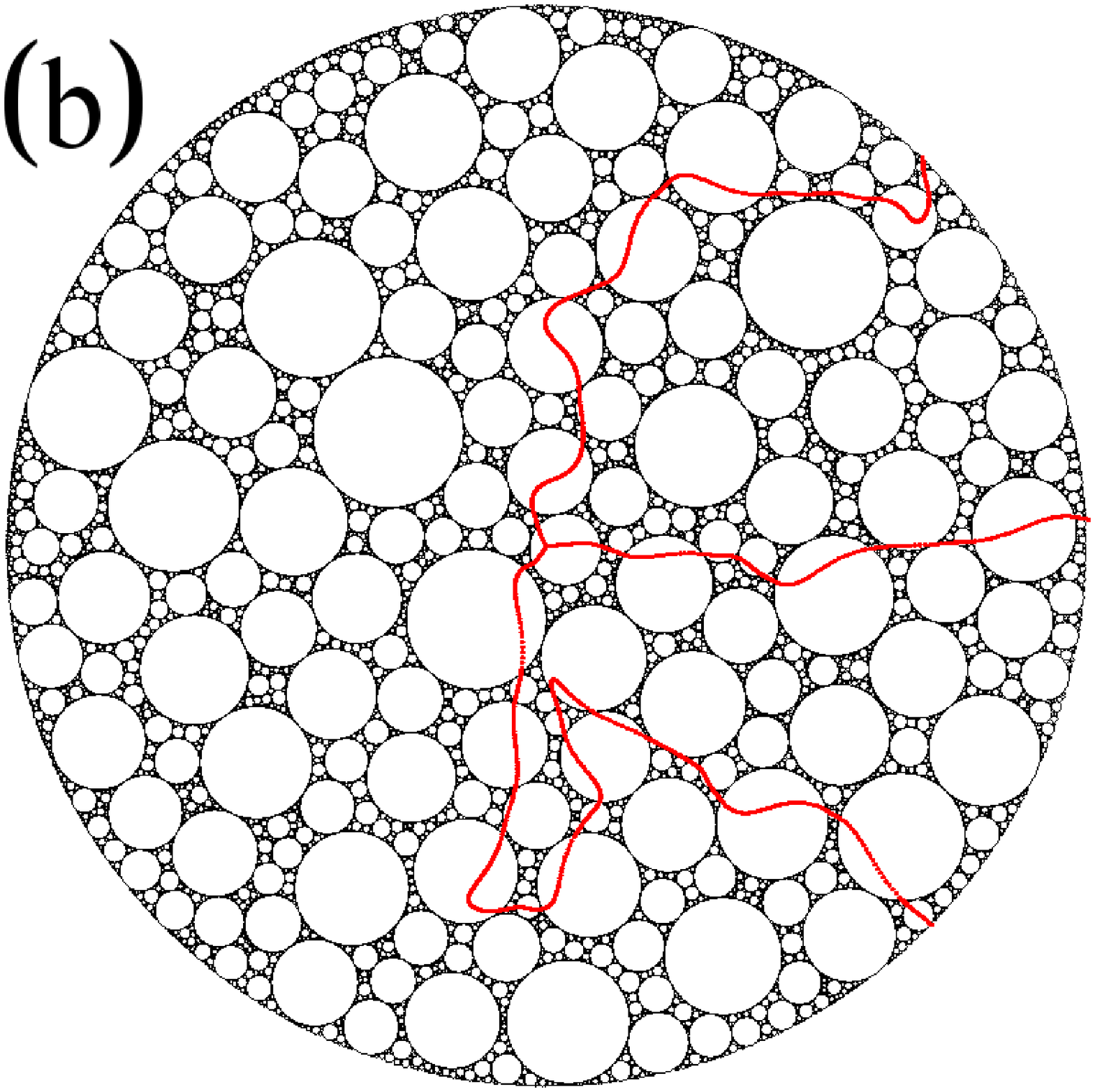}
\end{center}
\caption{%
(Color Online). {\bf (a)} Modelling multi-scale velocity fields,
using rigid discs to represent vortices (eddies) of different sizes.
Each eddy has a fixed center and rotates clockwise (green) or
counter-clockwise, in order to ensure absence of sliding at the
contact points. {\bf (b)} 
Trajectories of three massive particles
when particles follow Stokes drag (see Eq.~(\ref{eq:drag})).}
\label{fig1}
\end{figure}

\section{Bearings as models for velocity fields}

Our model for the velocity field of the flow is based on random
polydisperse packings of discs fulfilling the particular condition of
all discs being able to rotate simultaneously without sliding on each
other at their contact points. At their contact points all discs have
the same tangential velocity $v_0$ in order not to have frustration.
Therefore, $v_0$ is a global parameter of the system while the angular
velocity of disc $i$ is inversely proportional to its size,
$v_0=\omega_i R_i$. 

Figure \ref{fig1}a represents such a random
packing where colors indicate the sign of the angular velocity of the discs. 
One notes that no two discs which are in contact will have the 
same color since they rotate in opposite directions.  Such bichromatic
packings are referred to as bearings and were originally introduced in
the field of granular materials as a geometrical explanation of
seismic gaps in tectonic plates \cite{herrmannBearing1990,baram3dBearing2004,baramRandomBearing2005}. 

Random bearings used here are constructed by directly imposing the {\em
bichromacy} condition. Initially a number of discs of random sizes are
distributed randomly in the system. To each disc one color or the
other is assigned randomly. The rest of the space is filled by
inserting the largest possible discs with colors chosen such that no
two discs of the same color touch each other .

The configurations are complete (space-filling), self-similar and contain 
some level of randomness introduced by the initialization and the choice of 
parameters in the construction procedure (see Refs. \cite{lindBearing2008,baram3dBearing2004,baramRandomBearing2005} for more details.) 
Following Mandelbrot \cite{Mandelbrot} one can define the Hausdorff-dimension 
of the void space, which is related to the exponent of the size 
distribution of the discs.  
In practice, a lower cut-off on the size of the vortices is inevitable,
leading to a limited range of scales defined as $r_M/r_m$.

In this context, we regard each rotating disc as a vortex with a linear
velocity profile inside, that is, the fluid in each disc is moving in
the same way as if it were a rigid body. In this way, we obtain a
velocity field containing eddies of many different scales. 
Although, the statistical properties of such a flow field do not necessarily coincide with those of real 
turbulent flows, its energy spectrum follows a power-law whose exponent is 
directly related to the exponent of the size distribution of the eddies (see Ref.~\cite{herrmannBearing1990}).
During their rotation, the centers of the eddies are kept fixed which makes the velocity 
field stationary.

The particle-flow interaction depends on the shape of the particles
and the difference between their density and that of the fluid
\cite{dragMaxey1983}. Here, we will consider particles much denser
than the carrier fluid. Neglecting buoyancy and considering small
particle Reynolds numbers, when the size of the particles is much smaller than any active scale of the flow,  
the force exerted on the particles by the flow follows Stokes drag:
\begin{equation}
\frac{d\vec V_p}{dt}=\frac{\vec V_f-\vec V_p}{\tau},
\label{eq:drag}
\end{equation}
where $\tau$ is the response time of the particles and $V_f$ and $V_p$
are the velocities of the flow and the particle, respectively. For a
single vortex we have $\vec{V_f} = w\hat{z}\times(\vec{r}-\vec{r_c})$,
where $\vec{r}$ and $\vec r_c$ denote the position of the particle and
the center of the vortex, respectively, and $\hat{z}$ is the unit
vector perpendicular to the vortex. 
Equation (\ref{eq:drag}) is solved analytically to obtain particle trajectories.
Figure \ref{fig1}b shows three typical trajectories of particles through 
the system of eddies.

Inside the voids between eddies, we assume in Eq.~(\ref{eq:drag}) two different cases, namely, 
ballistic ($\vec{V_f}=\vec{V_p}$) and damping ($\vec{V_f}=0$).  
By assuming ballistic trajectories in the voids, we exclude all dissipation in the flow field. 
In the damped case the particles are slowed down which corresponds to a loss of energy and represents viscous dissipation on the scale of the smallest eddy, i.e.~ the Reynolds cut-off.

As the control parameter for flow-particle interaction, the Stokes number $S$ 
is often used instead of the response time $\tau$. It is dimensionless and related to $\tau$ through the following relation:
\begin{equation}
S=\frac{\tau}{\tau_c},
\label{eq:stokes}
\end{equation}
where $\tau_c$ is a characteristic time of the system. However, since
there is no unique characteristic time in our system we use $\tau$ in
our simulations for parameterizing the flow-particle interaction.

\begin{figure}
\begin{center}
\includegraphics*[width=0.24\textwidth]{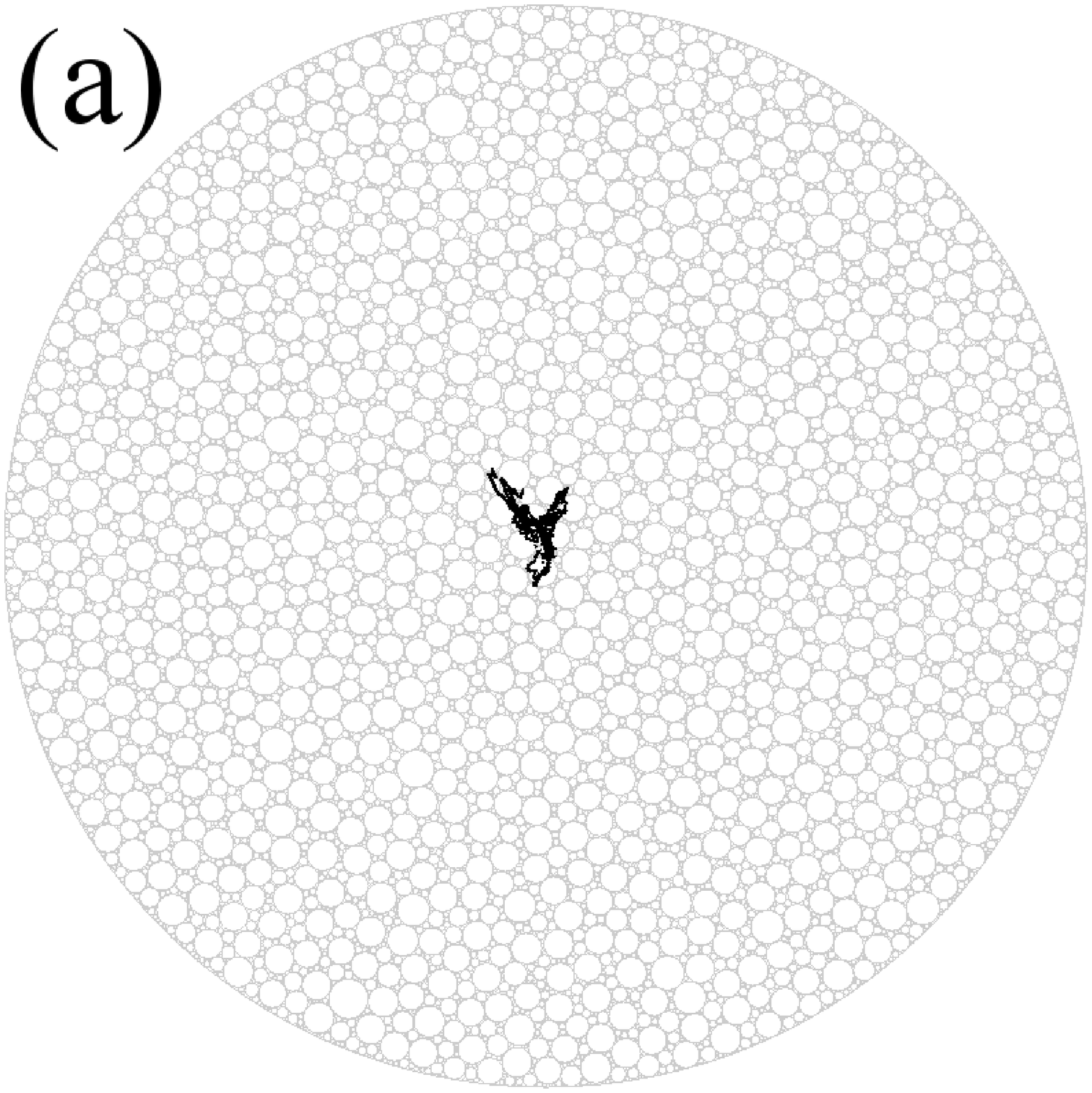}%
\includegraphics*[width=0.24\textwidth]{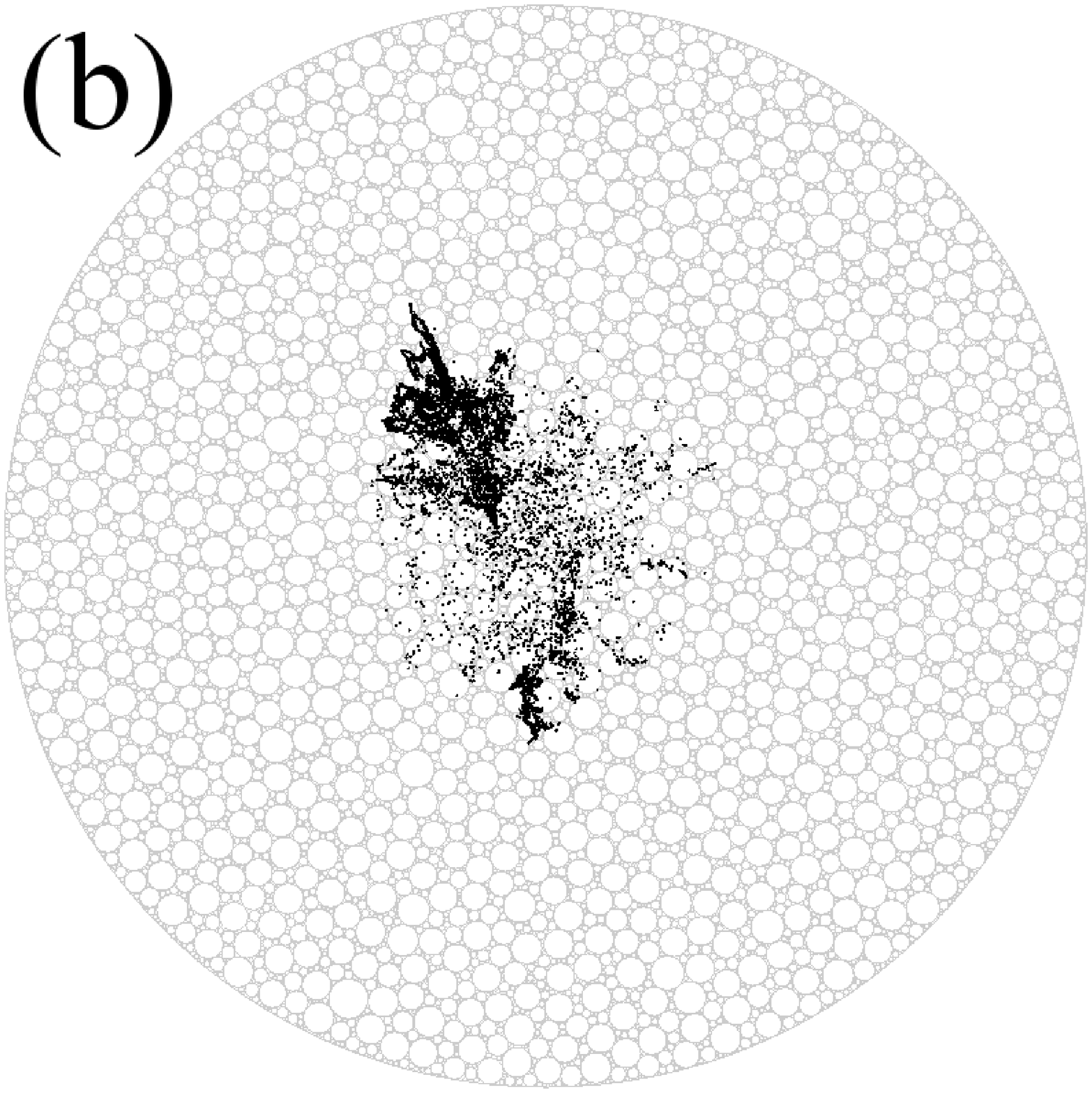}
\includegraphics*[width=0.24\textwidth]{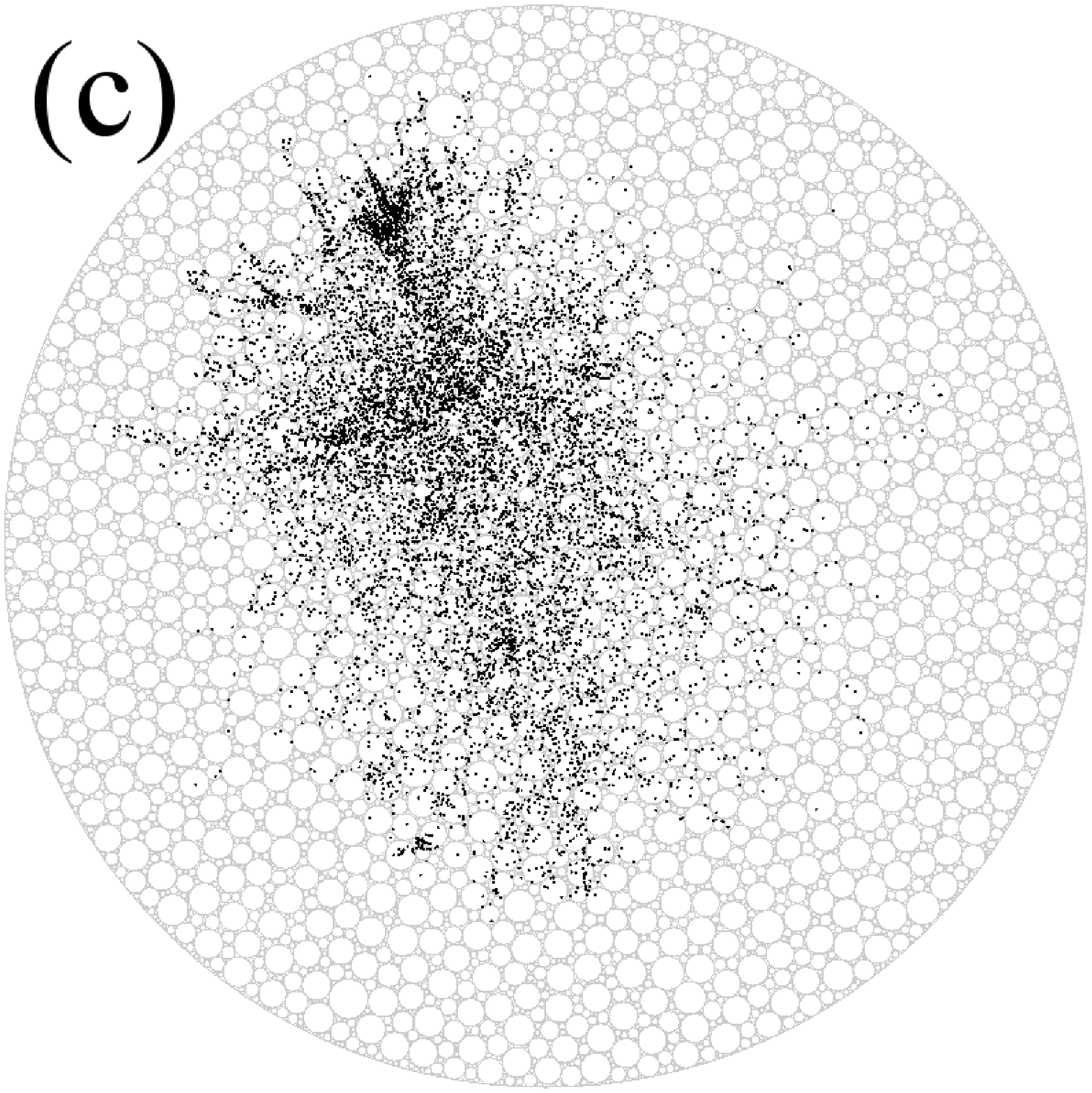}%
\includegraphics*[width=0.24\textwidth]{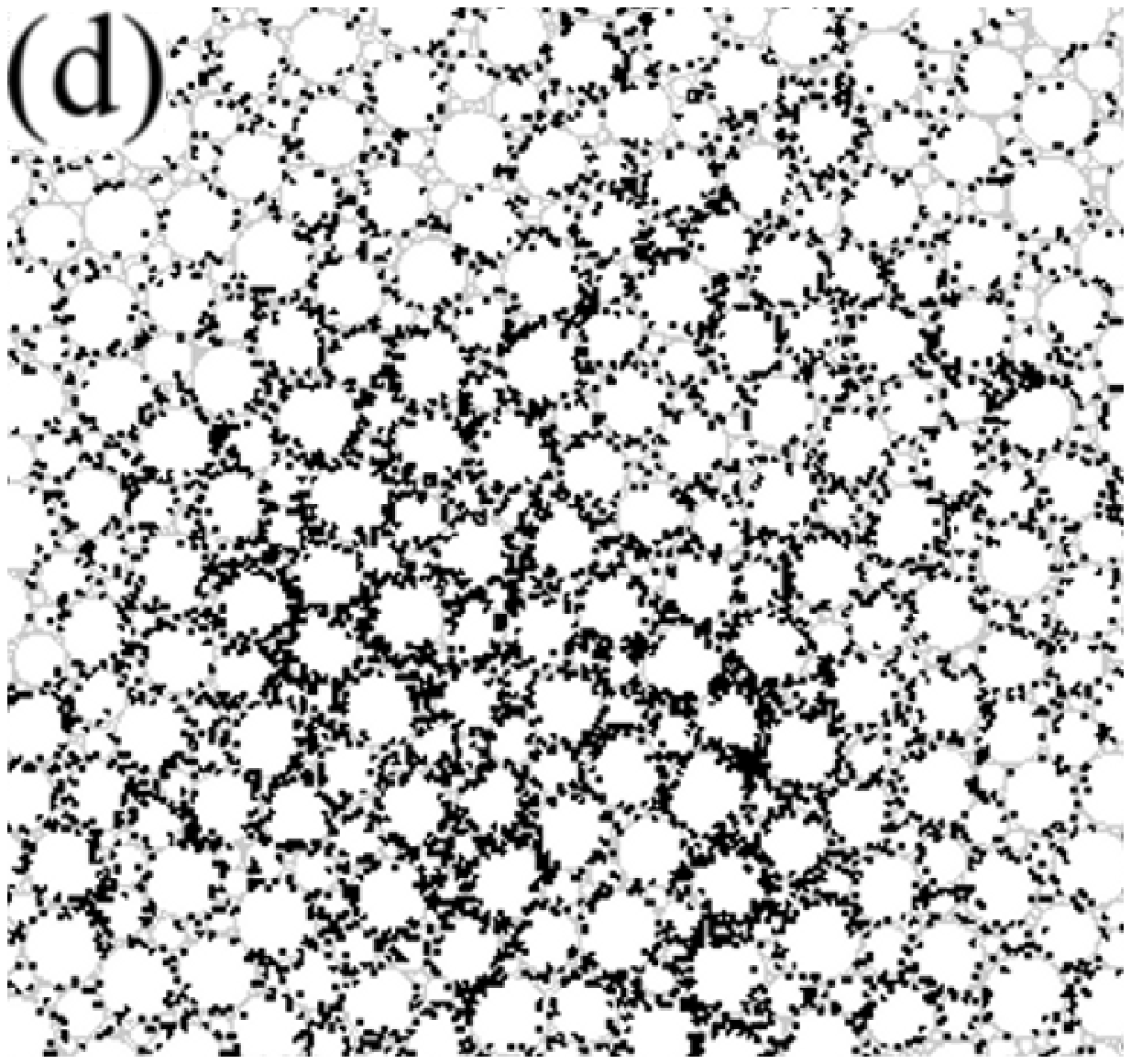}
\end{center}
\caption{\protect
Different stages during the diffusion of $10^4$ particles: {\bf (a)}
$\tau=14$, $t=2$, {\bf (b)} $\tau=14$, $t=6$, {\bf (c)} $\tau=14$,
$t=10$.  {\bf (d)} $\tau=0.01$, $t=3$.  The velocity field is defined
through $1.6\times 10^4$ vortices embedded in a circular space of unit
size (shown in the background) where the centers of the vortex stay
fixed in time. The size of the vortices varies from $r_M=0.02$ down to
$r_m=10^{-4}$ in units of system size ($r_M/r_m=200$).}
\label{fig2}
\end{figure}

\section{Superdiffusion and preferential concentration}
Having set up the velocity field as described above, a total number of
$10^4$ particles are released in a small region in the center of the system with very
small initial velocity in randomly chosen directions.
Figure \ref{fig2}a, \ref{fig2}b, and \ref{fig2}c represent,
respectively, the early, intermediate and final stages for one specific configuration with 
a moderate value of $\tau$. 
Figure \ref{fig2}d shows the final stage for a very small value of $\tau$ and here the preferential concentration becomes visible.

To improve the statistics, we average the results over $50$ different
eddy configurations, where the cut-off of the eddy size is
$r_m=10^{-4}$, and the maximum size is $r_M=0.02$ in units of the
system diameter.  In analogy to turbulence, the range of spatial
scales between the sizes of the smallest and the largest discs will be
here referred to as inertial range. 
In the simulations, the particles which reach the boundary of the system are removed permanently.
In order to prevent this from affecting the results, we gaurantee reaching the mixed state before a signifacant 
number of particles has left the system by choosing the size of the 
largest eddies to be much smaller than the system size.

\begin{figure}
\begin{center}
\includegraphics*[width=0.48\textwidth]{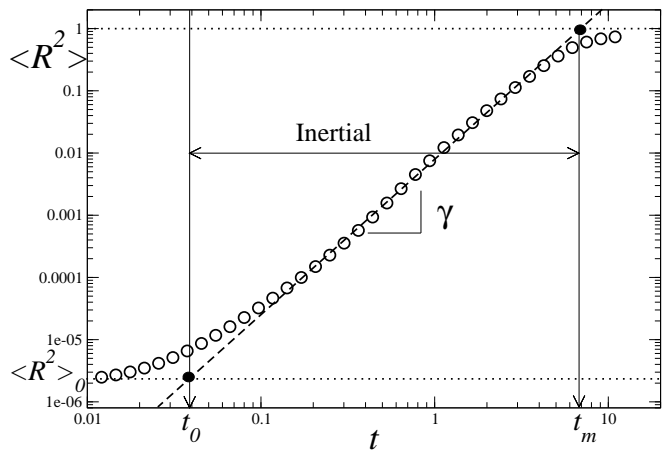}
\caption{
  Mean square separation $\langle R^2\rangle$ 
  between two particles as function of time $t$ 
 for a specific response time $\tau=3$. 
  Distances are in units of the system size $L$, the radius of the circular space, and time is in
  units of $L/v_0$, where $v_0$ is the tangent velocity of the discs.
  The two times $t_0$ and $t_m$ delimit the interval within
  which particles are in the inertial regime (see text).
  Time $t_0$ marks the beginning of the inertial regime whereas
   beyond $t_m$ particles are uniformly
  distributed in the system ($\langle R^2\rangle \sim 1$).
  The dashed line has slope $\gamma$ and its intersection
  with the dotted lines enables one to derive $t_0$ and $t_m$
  (see text).}
\label{fig3}
\end{center}
\end{figure}

Figure \ref{fig3} shows the mean square pair distance $\langle R^2
\rangle$ between two particles as function of time.
The initial regime (dotted line) for very short times, $t \le t_0$, 
corresponds to the period when the particles have not yet penetrated into 
the eddies of different scales.  
For large times, $t\ge t_m$, particles either escape or spread 
uniformly throughout the system yielding $\langle R^2\rangle \sim L^2$, 
as indicated by the horizontal dotted line, where $L=1$ is the radius of the system.
In the intermediate regimes, $t_0\le t \le t_m$, where 
particles are still in the inertial range, one observes 
$\langle R^2\rangle \sim t^{\gamma}$,
since these are regimes where the multi-scale structure of
the flow is dominant. These regimes are limited to times when the particles have not yet gone 
 through scales much larger than the size of the largest eddies. This behavior is qualitatively similar to that of observed 
 for separation of fluid particles via direct numerical simulations Ref.~\cite{boffetta2002}.

The values for $\gamma$, $t_0$ and $t_m$ depend on the response
time $\tau$, and are found using a power-law fit to $\langle R^2\rangle $ in the inertial range $[t_0, t_m]$, 
which can be written in the following form,
\begin{equation}
\langle R^2\rangle = \langle R^2\rangle_0 
       \left ( \frac{t}{t_0}
       \right )^{\gamma}
	\label{eqDiff}
\end{equation}
where $\langle R^2\rangle_0$ denotes the mean square separation of the particles
before entering the inertial range.
\begin{figure}
\begin{center}
\includegraphics*[angle=0, width=0.48\textwidth]{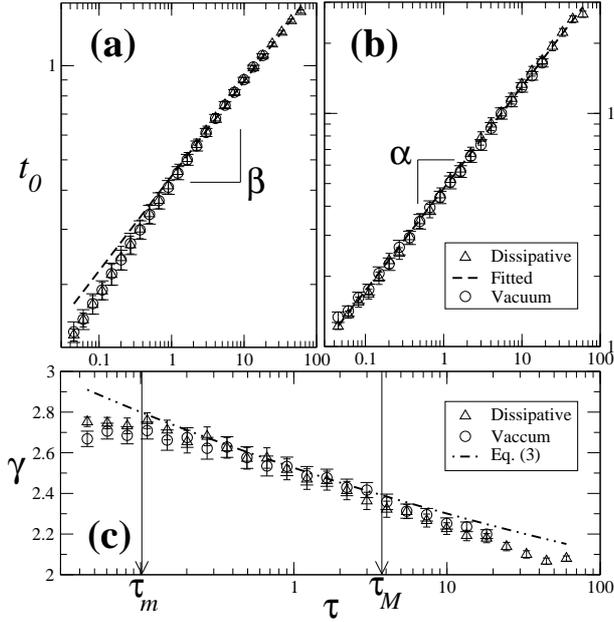}
\caption{\protect
   {\bf (a)} The transient time $t_0$ as a function of $\tau$ together with
   {\bf (b)} the mixing time $t_m$ and
   {\bf (c)} the exponent $\gamma$.
   Fits in (a) and (b) yield power-laws, namely $t_0\sim \tau^{\beta}$
   and $t_m\sim \tau^{\alpha}$ 
   (see text). In (c) the dashed line indicates $\gamma$ in 
   Eq.~(\ref{eqgamma}) where $a=-6$, $b=2.38$ and $\delta=0.101$.}
\label{fig4}
\end{center}
\end{figure}

As shown in Figs.~\ref{fig4}a and \ref{fig4}b, $t_0$ and 
$t_m$ behave as power-laws of $\tau$, namely $t_0= B\tau^{\beta}$ and 
$t_m= A\tau^{\alpha}$ respectively, independently of having dissipative
voids or vacuum. The fit beyond $\tau$ (dashed lines) yields,
$A=4.7\pm 1.2$, $\alpha=0.44\pm 0.14$, $B=0.437\pm 0.005$ and
$\beta=0.303\pm 0.005$. 
The value for
$\log{\langle R^2\rangle_0}=\gamma\log{(t_0/t_m)}\simeq 10^{-6}$
does not depend on $\tau$ but only on the specific initialization of the system.
Thus, from Eq.~(\ref{eqDiff}) we can write $\gamma$ as a function of $\tau$,
\begin{equation}
\gamma = \frac{a}{b+\delta\log{\tau}}
\label{eqgamma}
\end{equation}
with $a=\log{\langle R^2\rangle_0}$, $b=\log{(B/A)}$ and 
$\delta=\beta-\alpha$.
Figure \ref{fig4}c plots $\gamma$ as a function of $\tau$ (symbols) 
comparing it with Eq.~(\ref{eqgamma}) (dashed line) which fits the
observed values of $\gamma$ in the middle range of the $\tau$-spectrum, 
approximately at $\tau_1 < \tau < \tau_2$. The deviations from 
Eq.~(\ref{eqgamma}) imply that $t_0$ and $t_m$ obey a power-law only 
for the intermediate and not for the limiting values of $\tau$ which is what one should expect;
 (i) when $\tau\to\infty$, particles' inertia is very large and therefore they travel ballistically through the system without being 
significantly deflected by the flow, yielding $\gamma\to 2$, 
(ii) when $\tau\to 0$, which characterizes tracers (fluid particles), 
the motion of the particles follows Richardson's law $\langle R^2 \rangle \propto t^3$ \cite{richardson1926}.
Although the original Richardson's experiment was in three dimensions it can be
shown that Richardson's law is independent of the dimensionality of the system \cite{boffetta2002}.

These observations raise the hypothesis that the main ingredient
underlying Richardson's law is the multi-scale structure of vortices
with the dissipative dynamics playing a minor role. However, this cannot be 
rigorously verified due to the limitations of the model. As mentioned before, 
our velocity fields are stationary and lack the sweeping effect. Therefore, 
 the particles become trapped in local eddies in the limit of $\tau=0$ and cannot disperse. 

In the second part of our study, we provide an estimate of the range of 
$\tau$, $(\tau_1, \tau_2)$, showing that the effect of the limitations of the 
model is minor. This interval is also shown in Fig. \ref{fig4}c.

\begin{figure}[t]
\begin{center}
\includegraphics*[angle=0, width=0.48\textwidth]{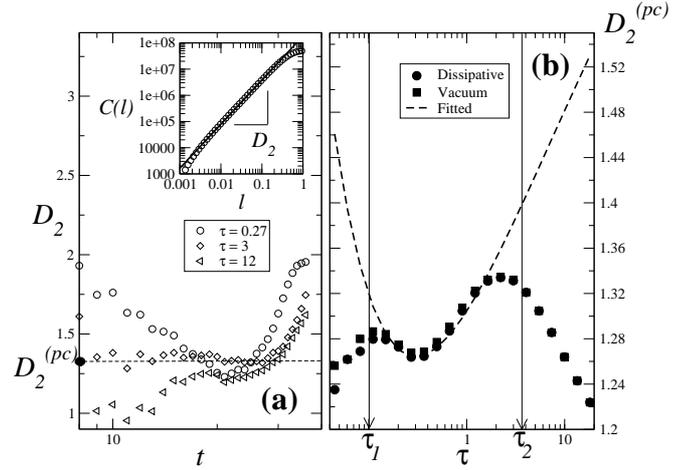}
\caption{\protect
     {\bf (a)} The correlation dimension $D_2$ as a function of
     time for three values of $\tau$.
     The inset shows how $D_2$ is determined: at each instant $t$,
     one computes the average number $C(l)$ of pairs of particles
     that are closer than a distance $l$; this yields a power-law 
     $C(l)\sim l^{D_2}$.
     At intermediate times $D_2$ attains a minimum (dashed line) which
     indicates maximal preferential concentration $D_2^{(pc)}$ (see text).
     {\bf (b)} The correlation dimension $D_2^{(pc)}$ when maximum 
     preferential concentration  is attained in the inertial range.
     The dashed curve fitting $D_2^{(pc)}$ in the range $\tau_1 < \tau < 
     \tau_2$  resembles the behavior found with DNS calculations of 
     particles in homogeneous turbulent flows\cite{bec07} (see text).}
\label{fig5}
\end{center}
\end{figure}

Let us next examine the occurrence and properties of preferential concentrations. 
To quantify the degree of preferential concentration we use the
correlation dimension as also used by other authors \cite{bec07},
defined as \cite{tangDispers1992, grassbergerStrange1983}:
\begin{equation}
   D_2 = \lim_{l\rightarrow 0} \frac{d(\log{C(l)})}{d(\log{l})},
\label{D2}
\end{equation}
where $C(l)$ is the fraction of pairs of particles separated by a
distance smaller than $l$. 
This quantity takes on the value zero, one and two when particles are all clustered in a single point,  
on a line, and uniformly distributed over the space, respectively. 
As can be seen from the inset of Fig.~\ref{fig5}a, in our case 
$C(l)\sim l^{D_2}$, which defines an effective exponent $D_2$ for each 
instant $t$.

Figure \ref{fig5}a shows the evolution of $D_2$ in time for
three values of $\tau$. For initial times, the correlation dimension 
depends strongly on the initial condition.
For very large times, one observes that $D_2$ converges to the
dimension of the system itself ($D_2=2$), i.e.~particles distribute
homogeneously throughout the system. 
For intermediate times, one observes a 
local minimum (dashed line), which implies that preferential concentration is maximal.  
For larger systems, we observe that the plateau where $D_2=D_2^{(pc)}$ 
becomes broader (not shown). Therefore, we regard $=D_2^{(pc)}$ as the characteristic 
value of the correlation dimension. While the results shown here are for the case of 
empty (ballistic) voids, similar results are obtained when the voids are dissipative. 

Figure \ref{fig5}b shows $D_2^{(pc)}$ as a function of $\tau$.  
Within the range $\tau_1 < \tau < \tau_2$ one finds a behavior
(dashed line) consistent to what is known for the correlation dimension
in numerical experiments of heavy particles in turbulent fluids, namely for $\tau=0$ and $\tau\to\infty$ the 
correlation dimension converges to the dimension of the system (see Ref.~\cite{bec07}).

The deviations from such behavior observed outside the range 
$\tau_1 < \tau < \tau_2$ are due to the finite range of scales of eddies.
As mentioned before, the particles are initialized inside a small neighborhood moving in a randomly chosen direction. 
For large $\tau$ values ($\tau>\tau_2$), particle paths are approximately 
ballistic, moving apart from each other as a growing circle at large 
times which yields $D_2\sim 1$. 
For $\tau<\tau_1$, particle are approximately tracers and get quickly 
trapped in a circular trajectory inside one vortex.
This last behavior is strengthened when voids are dissipative
(bullets), since particles are trapped in the voids.

\section{Conclusions}
We introduced a two-dimensional heuristic model for multi-scale
fluid velocity fields and studied the dependence of the drag-induced 
diffusion of massive particles on it.
The model is based on discs of different
sizes representing flow eddies which roll on each other without
sliding. By computing the exponent of the mean squared displacement as
a power-law of time we find superdiffusive regimes within a broad
range of response times which is consistent with scaling analysis
and with the particular limiting cases where Richardson's law holds
and also with the ballistic regime for heavy particles.
The results show that the main feature responsible for the
emergence of superdiffusion is the multi-scale vortical structure of
the flow. 
We also find the characteristic footprint of preferential concentration as observed when particle disperse in turbulent flows, like for instance the plankton in the ocean.
Being able to incorporate multi-scale features and at the same time
separate distinct mechanisms for heavy particle dynamics in fluids,
this model can be further extend to more general situations where,
e.g.~dissipation occurs inside the vortices.

\acknowledgments
RMB thanks {\it Funda\c{c}\~ao para a Ci\^encia e a Tecnologia} and PGL thanks  {\it Funda\c{c}\~ao para a Ci\^encia e a Tecnologia -- Ci\^encia 2007} for financial support.
  

\end{document}